\documentclass{ifacconf}

\usepackage{array}		
\usepackage{multirow}
\usepackage{xcolor}
\usepackage{float}
\usepackage{wrapfig}
\usepackage{enumitem}
\usepackage{fancyhdr}
\usepackage{amsmath}
\usepackage{amssymb}
\usepackage{mathtools}
\usepackage{cancel}

\newcommand{\nl}{ \\}
\newcommand{\dnl}{ \\[10pt]}
\newcommand{\mnl}{ \\[8pt]}

\usepackage{graphicx}      
\graphicspath{{./LatexPics/}}
\usepackage{natbib}        
\begin{document}
\begin{frontmatter}

\title{Tuning of CgLp based reset controllers: Application in precision positioning systems} 

\author[First]{Mahmoud Shirdast Bahnamiri}
\author[Second]{Nima Karbasizadeh}
\author[Second]{Ali Ahmadi Dastjerdi}
\author[Second]{Niranjan Saikumar} 
\author[Second]{S. Hassan HosseinNia}

\address[First]{Department of Mechanics and Mechatronics, University of Minho, 
   Guimaraes, Portugal (e-mail: m.shirdast@gmail.com).}
\address[Second]{Department of Precision and  Microsystems Engineering, Delft University of Technology,
   Delft, The Netherlands (e-mail: \{n.karbasizadehesfahani; a.ahmadidastjerdi; n.saikumar;
s.h.hosseinnia\}@tudelft.nl)}

\begin{abstract} 
This paper presents the tuning of a reset-based element called ``Constant in gain and Lead in phase" (CgLp) in order to achieve desired precision performance in tracking and steady state. CgLp has been recently introduced to overcome the inherent linear control limitation - the waterbed effect. The analysis of reset controllers including ones based on CgLp is mainly carried out in the frequency domain using describing function with the assumption that the relatively large magnitude of the first harmonic provides a good approximation. While this is true for several cases, the existence of higher-order harmonics in the output of these elements complicates their analysis and tuning in the control design process for high precision motion applications, where they cannot be neglected. While some numerical observation-based approaches have been considered in literature for the tuning of CgLp elements, a systematic approach based on the analysis of higher-order harmonics is found to be lacking. This paper analyzes the CgLp behaviour from the perspective of first as well as higher-order harmonics and presents simple relations between the tuning parameters and the gain-phase behaviour of all the harmonics, which can be used for better tuning of these elements. The presented relations are used for tuning a controller for a high-precision positioning stage and results used for validation.
\end{abstract}
\end{frontmatter}
\section{Introduction}
PID controllers continue to play a major role in the industry including precision motion applications like photolithography wafer scanners, atomic force microscopes due to their simplicity of design, implementation and compatibility with a wide range of applications. Additionally, PID controllers can be designed in the frequency domain using the well-known and studied loopshaping technique where the open-loop including controller and plant are shaped to achieve the required performance metrics in terms of tracking/disturbance rejection, stability/robustness and noise attenuation. However, all linear controllers suffer from the fundamental limitations of linear control represented by ``Bode's gain-phase relationship" and ``waterbed effect". This results in a trade-off between tracking precision and noise attenuation performance on one side and stability and robustness on the other side which has to be overcome if the ever increasing demands from the high-tech industry are to be met.

Among nonlinear control techniques, reset control has gained significant traction over the years due to its simple structure and compatibility with PID. \cite{clegg} introduced the reset integrator (henceforth referred to as Clegg integrator (CI)), where the integrator state value is reset to zero when the error input is equal to zero. Through describing function analysis, Clegg showed that CI had the same gain slope with a significantly reduced phase lag of only $38^\circ$ compared to $90^\circ$ of a linear integrator. This idea was extended several years later in the form of ``First order reset element" (FORE) by \cite{fore} which allowed for a first-order filter like design with the reset advantage. Apart from the tuning freedom, FORE was also used to show significant improvement in closed-loop control performance. Over the years, reset elements such as ``Second order reset element" (SORE) (see \cite{sore}) and ``Fractional order reset element" (FrORE) (see \cite{saikumar2017generalized}) have been introduced to provide greater flexibility in tuning of these reset elements, with tuning freedom also being provided through techniques such as partial reset (see \cite{banos2011reset}) resulting in generalized reset elements in \cite{gfore,cglp}.

The advantage of reset in feedback control is investigated in several works in literature from process to motion control systems (see \cite{banos2011reset,chen2001analysis,zheng2000experimental,hosseinnia2013fractional,beker2001plant,wu2007reset,reset,palanikumar2018no,chen2018beyond,chen2019development,akyuz2019reset,valerio2019reset,9030150}). While most of these works have mainly looked at the phase lag reduction advantage seen with reset, a combination of reset and linear filter for broadband phase compensation was introduced by \cite{cglp} and used for hysteresis compensation as part of a disturbance observer in \cite{saikumar2019resetting}. The describing function of this element showed unity gain with phase lead over a wide range of frequencies and hence was named ``Constant in gain Lead in phase" (CgLp). \cite{cglp} also used CgLp as part of the PID framework to show improvements in tracking and steady-state precision as well as improvement in bandwidth.

While significant advantages in performance improvement have been seen with reset and especially with CgLp, the use of describing function for performance prediction and tuning of these controllers has not always proved accurate and deviations from expected performance were noted in \cite{cglp,akyuz2019reset}. This is because describing function approximation in the frequency domain assumes that the first harmonic dominates the other harmonics. However, even when this is true, this approximation can prove insufficient for precision control applications. Hence the idea of describing functions was extended to include the higher-order harmonics by \cite{nuij2006higher} for nonlinear systems in general as higher-order sinusoidal input describing functions (HOSIDFs) and was applied for reset controllers by \cite{heinen2018frequency}. HOSIDFs provide a more complete representation of the reset controller behaviour in the frequency domain and have the potential to enable better tuning of these controllers. While the use of HOSIDFs for better tuning of CgLp was attempted by \cite{yusuf,hou}, no methods with a systematic analysis can be found in literature and this is the gap that this paper addresses.

The remainder of this paper is organized as follows. The required preliminaries for reset control are provided in Section 2. The describing functions and HOSIDFs are simplified for the reset elements which are used as part of CgLp in Section 3 and are analysed for tuning in Section 4. Based on the analysis,a simple design procedure is given in Section 5 which is validated with the results of implementation on a precision positioning stage in Section 6, followed by conclusions.

\section{Preliminaries}
The preliminaries related to reset control definition, describing functions, reset elements and CgLp are provided in this section.

\subsection{Reset controller definition}
While reset controllers with various state/input/time dependent laws exist in literature, the most popular reset law is based on the zero-crossing of the error input and this general SISO reset controller can be defined as follows:
\begin{align} \label{reset}
    &\Sigma_R =
    \begin{cases}
        \dot{x}(t) =A x(t) + B e(t) & \ e(t)\neq 0 \nl
        x(t^+) =A_\rho x(t)& \ e(t)=0 \nl
        u(t) =C x(t) + D e(t)\nl
    \end{cases}
\end{align}
where $e(t) \in $ is the error input, $u(t)$ is the controller output and $x(t) \in \mathbb{R}^{n_r}$ are the states of the controller. $A$, $B$, $C$, and $D$ represent the state-space matrices and are together referred to as the base-linear controller. The first equation provides the non-reset continuous dynamics referred to as flow dynamics, whereas the resetting action is given by the second equation referred to as the jump dynamic. $A_\rho$ is resetting matrix which determines the after-reset values of the states and is generally of form $diag(\gamma_1,\gamma_2,....,\gamma_{n_r})$ where $\gamma_i \in [-1,1]$. 

\subsection{Describing function (DF)} \label{sec.df}
Due to the nonlinear nature of the controller, sinusoidal input describing function (DF) is used for frequency domain analysis. However, an isolated use of DF neglecting the higher-order harmonics is insufficient when designing controllers for high precision applications \cite{akyuz2019reset,9030150}. Hence the idea of higher-order sinusoidal input describing function (HOSIDF) presented in \cite{nuij2006higher} was extended for reset controllers by \cite{heinen2018frequency}. The DF and HOSIDFs for a reset controller defined in (\ref{reset}) can hence be obtained analytically as
\begin{equation}
	G_n(\omega) =
	\begin{cases} \label{df}
		C (j\omega I-A)^{-1} (I+j\Theta_D (\omega))  B+D & n=1 \nl
		C (jn\omega I-A)^{-1} (j\Theta_D (\omega))  B & \text{odd} \ n>1  \nl
		0 & \text{even} \ n>1 
	\end{cases}
\end{equation}
where $n$ is the order of the harmonic and
 \begin{eqnarray*}
 	\left.\begin{aligned}
	&\Lambda(\omega) = \omega^2 I+A^2\\
	&\Delta(\omega) = I+e^{(\tfrac{\pi}{\omega} A)}\\
	&\Delta_r (\omega) = I+A_\rho e^{(\tfrac{\pi}{\omega} A)}\\
	&\Gamma_r (\omega) = \Delta_r^{-1} (\omega)  A_\rho  \Delta(\omega)  \Lambda^{-1} (\omega) \\
	&\Theta_D (\omega) = \dfrac{-2\omega^2}{\pi}  \Delta(\omega)  \left[ \Gamma_r (\omega)- \Lambda^{-1} (\omega)\right]
	\end{aligned}\right.
\end{eqnarray*}

\subsection{Stability of Reset control systems}

The closed-loop stability of systems consisting of reset controllers defined as in (\ref{reset}) in the feedback loop has been extensively studied in literature.  We refer the readers to the work of \cite{beker2004fundamental} for $H_\beta$ conditions and also to the work of \cite{nevsic2008stability} using piece-wise Lyapunov equations. We assume that stability checks are automatically part of the design procedure irrespective of the chosen stability condition and is not explicitly mentioned henceforth.

\subsection{Reset elements}
While several reset elements exist in literature, we present the relevant ones here.

\subsubsection{GFORE: } FORE presented by \cite{fore} was generalized and extended as GFORE by \cite{gfore} allowing for a first-order filter like reset element with control over the resetting matrix $A_\rho$. A GFORE element with its corner frequency at $\omega_r$ can be represented as below with the arrow indicating reset.
\begin{align} \label{fore}
	\text{GFORE}= \dfrac{1}{\cancelto{A_\rho}{\frac{s}{\alpha \omega_r} +1}}
\end{align}
where $\alpha$ accounts for the change in the gain of GFORE at high frequencies as noted in \cite{cglp}, $A_\rho = \gamma \in [-1,1]$ with the value of $\alpha$ depending on the value of $\gamma$. The corresponding state-space matrices as per (\ref{reset}) are given as
\begin{equation}
	A = -\alpha\omega_r,\quad B=\alpha\omega_r,\quad C=1,\quad D=0 \nonumber
\end{equation}

\subsubsection{GSORE: } SORE allows for additional tuning of the damping parameter of the filter and is the reset equivalent of a linear second order low-pass filter. SORE presented in \cite{sore} was generalized in \cite{cglp} and can be represented as:
\begin{equation} \label{sore}
	\text{GSORE}= 
	\dfrac{1}{\Big(\cancelto{A_\rho}{\frac{s}{\kappa \omega_r}\Big)^2 + 2\beta \frac{s}{\omega_r}+1}}
\end{equation}
where $\kappa$ again corrects for the change in gain with $\omega_r$ being the corner frequency, $\beta$ being the damping coefficient and resetting matrix $A_\rho = \gamma I^{2\times2}$ with $\gamma \in [-1,1]$. The corresponding state-space matrices as per (\ref{reset}) are given as
\begin{align*}
	A =
	\begin{bmatrix} 0 & 1 \mnl -(\kappa \omega_r)^2 & - 2\beta \kappa^2 \omega_r \end{bmatrix}, \
	& B =
	\begin{bmatrix} 0   \mnl  (\kappa \omega_r)^2 \end{bmatrix}, \\
	C =
	\begin{bmatrix} 1 &  0 \end{bmatrix}, \
	& D = 0
\end{align*}

\subsubsection{CgLp:} ``Constant in gain Lead in phase" element is designed to obtain phase lead with unity gain over a broad range of frequencies by combining a reset element (GFORE or GSORE) with a corresponding order linear lead filter. The linear lead filter is of the form
\begin{equation} \label{linearlead}
L(s) = \dfrac{{\frac{s}{\omega_r} +1}}{{\frac{s}{\omega_f} +1}} \text{ or } \dfrac{\Big({\frac{s}{\omega_r}\Big)^2 + 2\zeta \frac{s}{\omega_r}+1}}{\Big({\frac{s}{\omega_f}\Big)^2 + 2\frac{s}{\omega_f}+1}}
\end{equation}
where $\omega_f >> \omega_r$ and $\zeta$ is the damping factor of the second order lead. If a linear lag and lead filter with the same corner frequency are placed in series, then they cancel each-other in both gain and phase. However, since a reset filter GFORE or GSORE have significantly less phase lag while retaining the gain behaviour of their linear counterparts, the combination resulting in CgLp provides unity gain with a corresponding phase lead in the frequency range $(\omega_r, \omega_f)$.

\section{simplification of describing functions}
The describing function of reset elements which approximates their behaviour in the frequency domain is accurate when the first harmonic dominates the other higher-order harmonics. Hence, it follows that if we are to use DF for loopshaping, then a reduction in the magnitude of the higher-order harmonics should allow for deviations between predicted and achieved performance to reduce. In this respect, the describing function equations are simplified in this section for different regions of frequency domain to allow for a simplified systematic analysis of the magnitude and phase behaviour of all the harmonics with the intention to look for tuning conditions allowing for better tuning of these elements. In this case, better tuning translates to ensuring required gain and phase behaviour are achieved by the first harmonic while the magnitude of higher-order harmonics are reduced to the maximum extent possible.

All systems considered in this section are assumed to be Schur stable. Hence for the reset controller defined as in (\ref{reset}) and DF and HOSIDF provided as in (\ref{df}), the simplifications at low and high frequencies are provided as follows. In the case of high frequencies, the simplification is only performed for reset controllers of order $n_r \leq 2$. For low frequencies,
\begin{align}
	&DF(\omega, n)_{lf}\approx
	\begin{cases}  \label{df_w_0}
		-C A^{-1} B+D & n=1  \nl
		j\dfrac{-2\omega^2}{\pi}(1-\gamma) C A^{-3}  B & \text{odd } \ n>1  \\
		0 & \text{even} \ n>1
	\end{cases}
\end{align}
For $n_r = 1$ at high frequencies,
\begin{align}
	&DF(\omega, n)_{hf} \approx
	\begin{cases} 
	C\dfrac{1}{j\omega}\left(1+j F\right)B+D  & n=1  \dnl
	C\dfrac{1}{jn\omega}\left(j F\right)B & 			\text{odd} \ n>1  \dnl
	0 & \text{even} \ n>1
	\end{cases}
\end{align}
where $F=\dfrac{4}{\pi} \cdot \dfrac{1-\gamma}{1+\gamma} \nonumber$. Subscript lf refers to low frequencies, i.e., small values of $\omega$ but not tending to zero, and subscript hf refers to high frequencies, i.e., large values of $\omega$ but not tending to infinity.

For $n_r = 2$ at high frequencies,
\begin{align}
	&DF(\omega, 1)_{hf} \approx \nonumber \\
	&C
	\begin{bmatrix} \dfrac{1}{ j\omega} &  \dfrac{-A_{12}}{ \omega ^2} \mnl  \dfrac{-A_{21}}{ \omega ^2}  &\dfrac{1}{j\omega} \end{bmatrix} 
	\begin{bmatrix} 1+j F& j F \   \dfrac{-A^2_{12}}{\omega^2} \mnl j F \  \dfrac{-A^2_{21}}{\omega^2}  & 1+j F\end{bmatrix}
	B+D \label{df_inf} \dnl
	&DF(\omega, n)_{hf}|_{\text{odd } n>1}\approx \nonumber \\
	&C
	\begin{bmatrix} \dfrac{1}{ jn\omega} &  \dfrac{-A_{12}}{ n^2\omega ^2} \mnl \dfrac{-A_{21}}{ n^2\omega ^2}  & \dfrac{1}{j n\omega} \end{bmatrix}
	\begin{bmatrix} j F &          j F \       \dfrac{-A^2_{12}}{\omega^2} \mnl j F \  \dfrac{-A^2_{21}}{\omega^2}  & j F\end{bmatrix}
	B  \label{hosidf_inf} 
\end{align}
where $A_{nm}$ and $A^2_{nm}$ represent the element of $n^{th}$ row and  $ m^{th}$ column in Matrices $A$ and  $A^2$, respectively.

Note: Due to errors in approximation of the term $e^{(\tfrac{\pi}{\omega} A)}$ at high frequencies, (\ref{df_inf})  and  (\ref{hosidf_inf}) are only precise enough to comprehend the changing trend of the reset system.

\subsection{Simplified describing functions to GFORE}
The simplified equations as applied to GFORE are presented as follows.
\begin{align}
    &\text{GFORE}(\omega,1)_{lf} \approx 1 \label{fore_m_zero}\\
    &|\text{GFORE}(\omega,n)_{lf}|_{\text{odd } n>1} \approx= \dfrac{2(1-\gamma)}{\pi} \  \dfrac{\omega^2}{(\alpha \omega_r)^2} \\
    &|\text{GFORE}(\omega,1)_{hf}|\approx \sqrt{1+F^2 \ } \ \dfrac{\alpha \omega_r}{\omega} \\
    &\angle \text{GFORE}(\omega,1)_{hf} \approx -\dfrac{\pi}{2}+\tan^{-1}(F)  \\
    &|\text{GFORE}(\omega,n)_{hf}|_{\text{odd } n>1 } \approx F \dfrac{\alpha \omega_r}{n\omega}
\end{align}

\subsection{Simplified describing functions to GSORE}
Similarly for GSORE, the equations of DF and HOSIDF can be simplified at low and high frequencies as
\begin{align}
&\text{GSORE}(\omega,1)_{lf} \approx 1 \label{sore_m_zero}\\
    &|\text{GSORE}(\omega,n)_{lf}|_{\text{odd } n>1 } \approx
    \left| \dfrac{2(1-\gamma)}{\pi} \ \dfrac{4\kappa^2\beta^2-1}  {(\kappa\omega_r)^2}\omega^2 \right| \label{sore_hosidf_0}\\
    &|\text{GSORE}(\omega, 1)_{hf}| \approx \sqrt{1+F^2} \ (\dfrac{\kappa \omega_r }{ \omega } )^2  \\
    &\angle \text{GSORE}(\omega, 1)_{hf} \approx \tan^{-1}(F) \\
    &|\text{GSORE}(\omega, n)_{hf}|_{\text{odd}  \ n>1} \approx (\dfrac{\kappa \omega_r }{ n\omega } )^2 \ F 
\end{align}

\begin{figure}
	\centering
	\includegraphics[width = \linewidth]{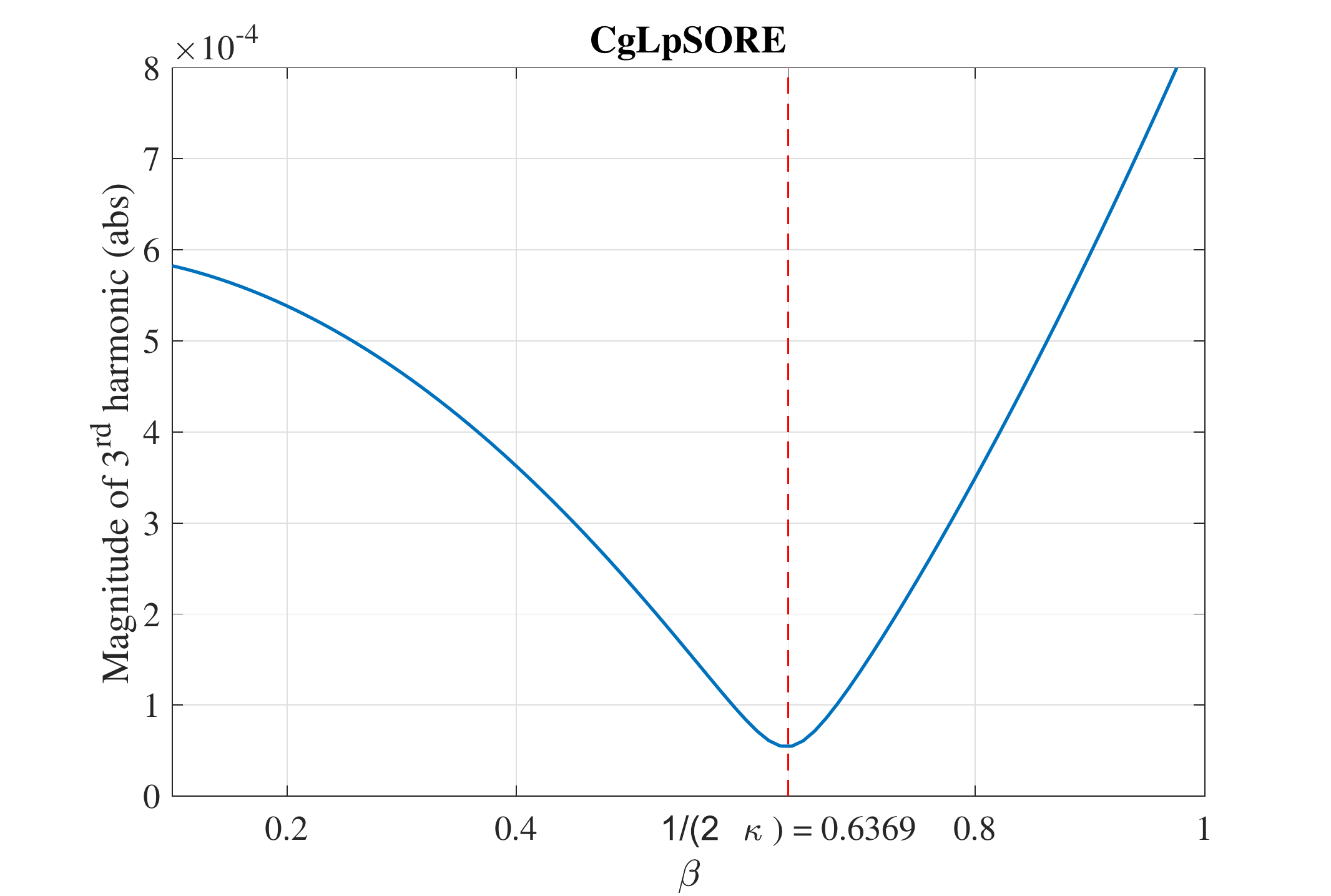} 
	\caption{Magnitude of $3^{rd}$ order harmonic of GSORE as a function of $\beta$ at $\omega= 5\ rad/sec$ with $\kappa=0.785$ determined for $\gamma = 0$ to correct for change in corner frequency (see \cite{cglp})}
	\label{sore_hosidf_zero_beta}
\end{figure}

According to (\ref{sore_hosidf_0}), choosing $\beta = \dfrac{1}{2\kappa}$ results in zero higher-order harmonics at low frequencies. Figure \ref{sore_hosidf_zero_beta} shows the trend of the magnitude of $3^{rd}$ order harmonic of GSORE versus the parameter $\beta$ at 5 rad/s, confirming the trend as expected from the simplified equations.

\subsection{CgLp}
CgLp is created through a series combination of GFORE or GSORE with a corresponding order lead filter. For the lead filters defined in (\ref{linearlead}), the magnitude and phase at low frequencies can similarly be simplified assuming $\omega_f \rightarrow \infty$ as follows. For the first order lead filter, we get,
\begin{align}
	&\text{L}(j\omega)_{lf} \approx 1 \label{nrfore_m_zero}\\
	&\text{L}(j\omega)_{hf} \approx \dfrac{\omega_r}{j\omega}
\end{align}
and in the second order case, we get,
\begin{align}
	&\text{L}(j\omega)_{lf} \approx 1 \label{nrsore_m_zero}\\
	&\text{L}(j\omega)_{hf} \approx \Big(\dfrac{\omega_r}{j\omega}\Big)^2
\end{align}
Hence, the simplification of the DF and HOSIDF equations for CgLp based on GFORE as well as that based on GSORE can be obtained as follows:
\begin{align}
	&|\text{CgLp-FORE}(\omega,n)_{lf}|_{\text{odd} \ n>1} \approx  \dfrac{2(1-\gamma)}{\pi} \ \dfrac{\omega^2}{(\alpha \omega_r)^2} \label{cglpfore_zero_hosidf_gain} \\
    &|\text{CgLp-FORE}(\omega,1)_{hf}| \approx \alpha \ \sqrt{1+F^2 \ }  \label{cglpfore_inf_df_gain}\\
    &\angle \text{CgLp-FORE}(\omega,1)_{hf} \approx \tan^{-1}(F) \label{cglpfore_inf_df_phase}\\
    &|\text{CgLp-FORE}(\omega,n)_{hf}|_{\text{odd} \ n>1 } \approx \alpha \ F \label{cglpfore_inf_hosidf_gain}
    \end{align}
    \begin{align}
    &|\text{CgLp-SORE}(\omega,n)_{lf}|_{\text{odd} \ n>1} \approx
    \left| \dfrac{2(1-\gamma)}{\pi} \ \dfrac{4\kappa^2\beta^2-1}  {(\kappa\omega_r)^2}\omega^2 \right| \label{cglpsore_zero_hosidf_gain}\\
    &|\text{CgLp-SORE}(\omega, 1)_{hf}| \approx \kappa^2  \ \sqrt{1+F^2} \label{cglpsore_inf_df_gain}\\
    &\angle \text{CgLp-SORE}(\omega, 1)_{hf} \approx \pi+\tan^{-1}(F) \label{cglpsore_inf_df_phase}\\
    &|\text{CgLp-SORE}(\omega, n)_{hf}|_{\text{odd} \ n>1} \approx \kappa^2 \ F \label{cglpsore_inf_hosidf_gain}
\end{align}

These simplified equations are used in the next section for analysis of CgLp elements.

\section{CgLp tuning} \label{cglp_tuning}
As noted earlier, the aim is to tune the CgLp element such that the required gain and phase behaviour of the first harmonic is achieved as accurately as possible to obtain the prescribed open-loop shape while at the same time reducing the magnitude of the higher-order harmonics to ensure reliability of using DF for loopshaping. The following analysis is based on (\ref{cglpfore_inf_df_gain}) to (\ref{cglpsore_zero_hosidf_gain}).
\subsection{Gain of $1^{st}$ harmonic}
\begin{figure}
\begin{center}
	\includegraphics[width = \linewidth]{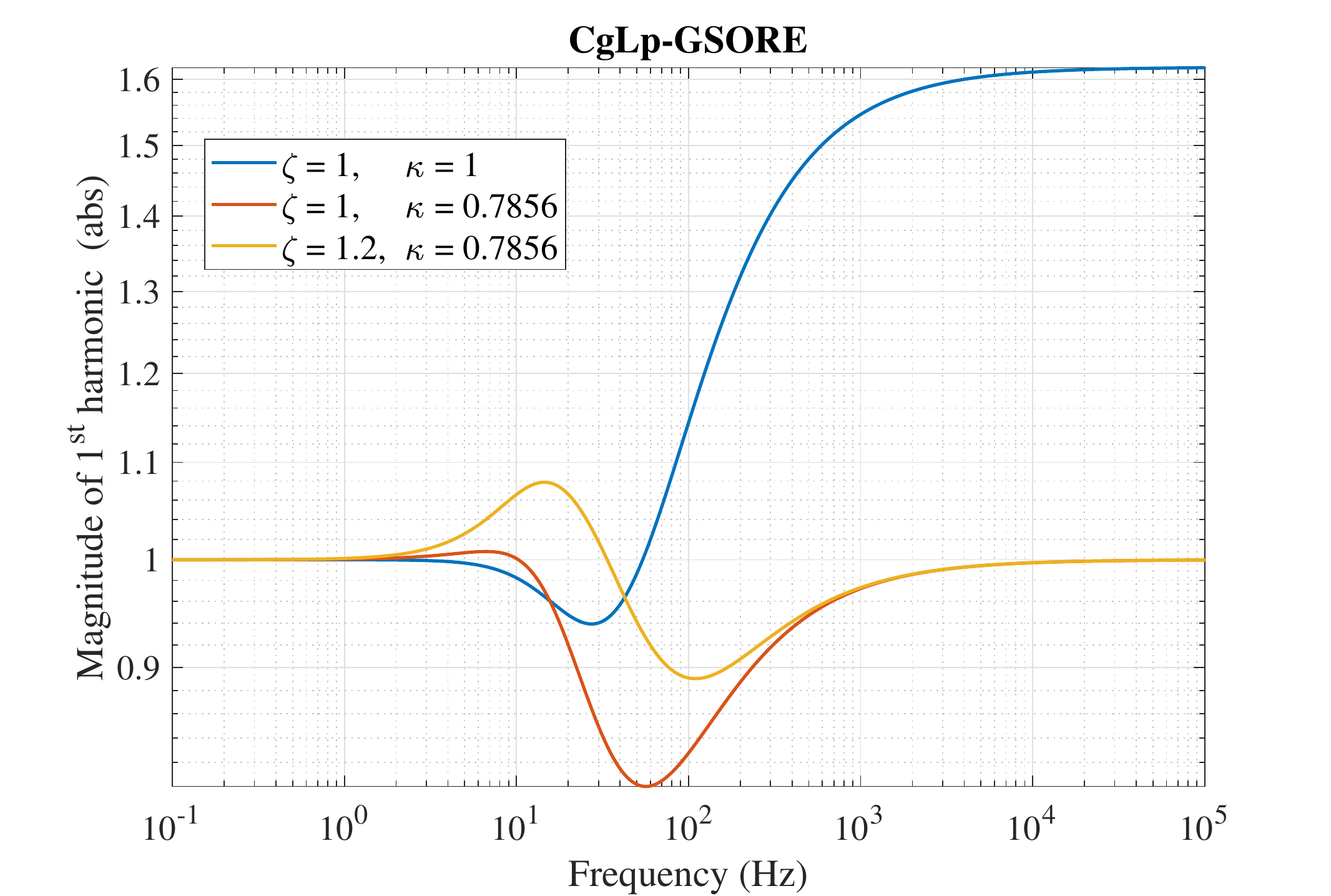} 
	\caption{Influence of choice of $\kappa$ and $\zeta$ on first order harmonic of CgLp-SORE at high and mid-range frequencies. ($\gamma=0$, $\beta = 1$, $\frac{1}{\sqrt[4]{1+F^2}} = 0.7856$)}
	\label{fig_cglpsore_gain}
\end{center}
\end{figure}

CgLp aims to obtain unity gain over the entire frequency range $(\omega_r,\omega_f)$. If as before, we assume $\omega_f \rightarrow \infty$, then this translates to a unity gain over the entire frequency range. From (\ref{fore_m_zero}), (\ref{nrfore_m_zero}), (\ref{sore_m_zero}), and  (\ref{nrsore_m_zero}), this is automatically achieved at low frequencies. However, at high frequencies, choosing the corrective parameters $\alpha$ and $\kappa$ as calculated in (\ref{alphachoice}) and (\ref{kappachoice}) results in unity gain according to (\ref{cglpfore_inf_df_gain}) and (\ref{cglpsore_inf_df_gain}).
\begin{align}
	\alpha=\dfrac{1}{\sqrt{1+F^2}} \label{alphachoice}\\
	\kappa=\dfrac{1}{\sqrt[4]{1+F^2}} \label{kappachoice}
\end{align}
Although, this choice ensures that unity gain is achieved at low and high-frequencies, the same cannot be achieved at frequencies close to $\omega_r$. In the case of CgLp-SORE, since $\beta = \dfrac{1}{2\kappa}$ is chosen to ensure reduced higher-order harmonics, the damping factor $\zeta$ can be chosen to ensure that minimum deviation from unity gain is achieved at all frequencies. Figure \ref{fig_cglpsore_gain} shows the influence of the choice of both $\kappa$ and $\zeta$ in CgLp-SORE in achieving unity gain over the entire frequency range.
\subsection{Phase lead of $1^{st}$ harmonic}
\begin{figure}
\begin{center}
	\includegraphics[width = \linewidth]{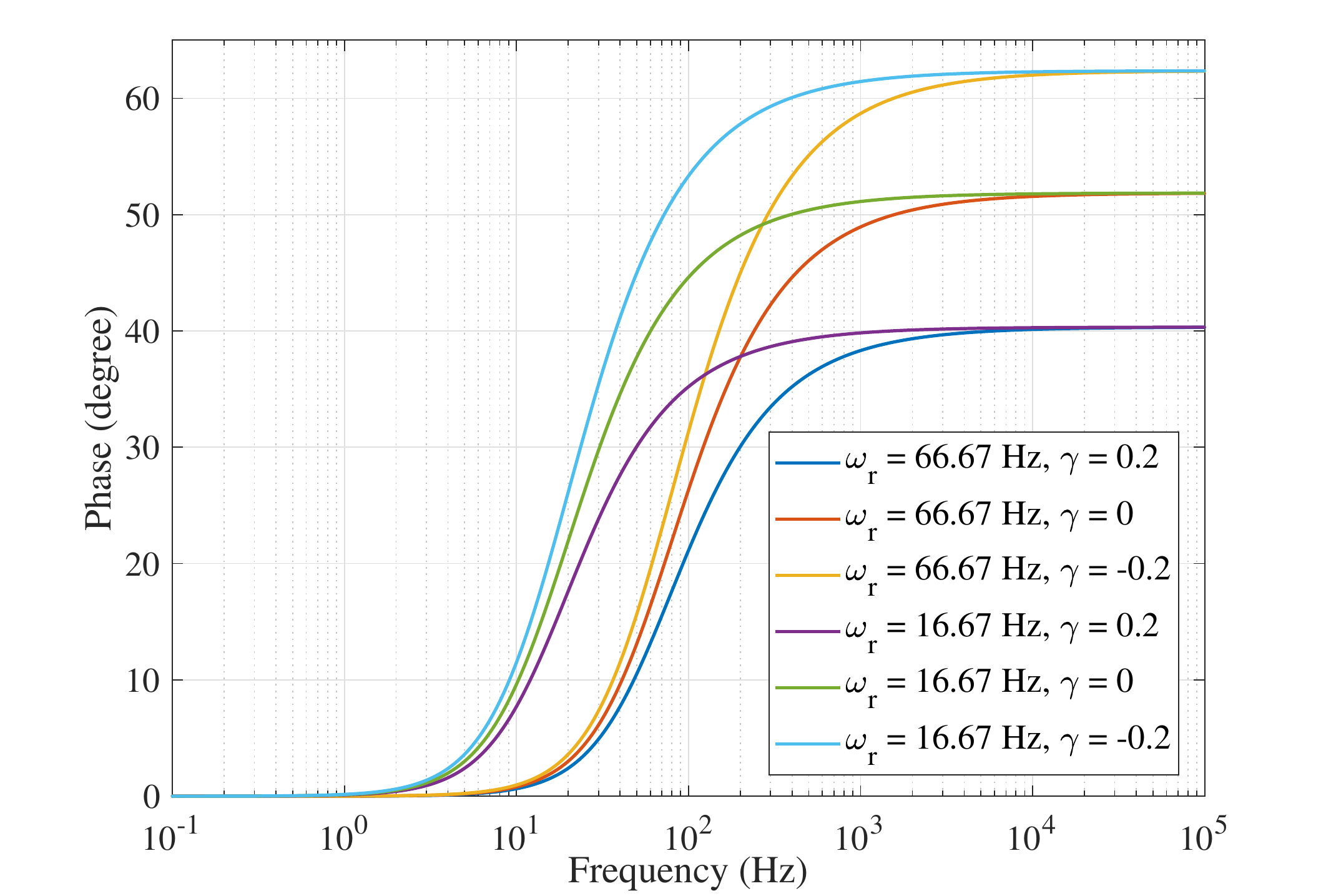} 
	\caption{Effect of choice of $\gamma$ and $\omega_r$ on the phase lead achieved with CgLp-FORE at the required bandwidth frequency ($100\ Hz$ in this case). $\alpha$ is chosen to ensure unity gain.}
	\label{fig_cglp_phase}
\end{center}
\end{figure}

\begin{figure}
\begin{center}
	\includegraphics[width = \linewidth]{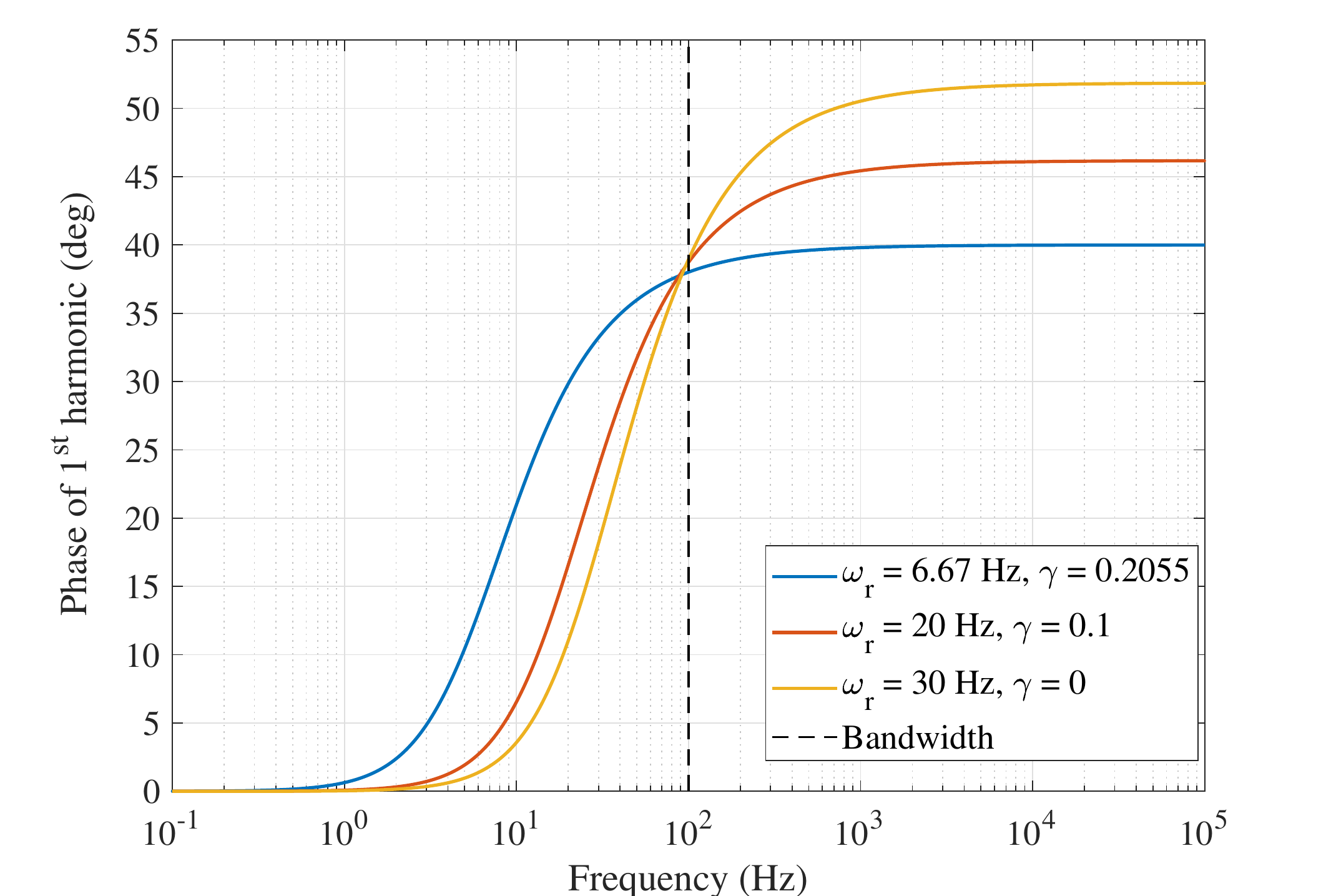} 
	\caption{Three different combinations of $\omega_r$ and $\gamma$ which produce $40^\circ$ of phase lead at $100\ Hz$. $\alpha$ is chosen to ensure unity gain.}
	\label{pm_gamma_wr}
\end{center}
\end{figure}

The phase lead of the CgLp as seen in its first harmonic arises from the fact that the reset element has lesser phase lag compared to the phase lead achieved by the linear lead filter. At low frequencies, CgLp has $0^\circ$ phase. The phase of CgLp starts rising before $\omega_r$ and from (\ref{cglpfore_inf_df_phase}) and (\ref{cglpsore_inf_df_phase}), it is clear that the asymptotic phase lead at high frequencies achieved by CgLp is only dependent on the value of $\gamma$. It is self-evident that in the context of loopshaping, that the value of $\omega_r$ should be appropriately chosen to be below or close to the bandwidth to obtain the required phase lead. From this it follows, that the phase lead obtained from CgLp at bandwidth is a function of $\omega_r$ and $\gamma$ (see \cite{cglp}). Figure \ref{fig_cglp_phase} demonstrates the effect of the choice of these two parameters on the phase lead achieved at the frequency of $100\ Hz$. This clearly shows that CgLp can either be tuned to have $\omega_r$ very close to the bandwidth with a low value of $\gamma$ or inversely with $\omega_r$ further to the left of bandwidth with a higher value of $\gamma$. This is more clearly shown in Figure \ref{pm_gamma_wr}, where the required phase lead is achieved with three different combinations of $\omega_r$ and $\gamma$. These chosen combinations however have different higher-order harmonic behaviour which is discussed next.

\subsection{Higher-order harmonics}
The idea of shaping the open-loop using DF is valid and more accurate when the first harmonic dominates the higher-order harmonics. Hence, it is the aim to reduce the magnitude of the higher-order harmonics at all frequencies of interest. Low frequencies are indeed of interest since they correspond to tracking region, while frequencies after the bandwidth are associated with noise attenuation. In real life implementation, due to sampling and discretization of controllers, any noise present at very high frequencies are not detectable according to Nyquist and hence the behaviour of higher-order harmonics is not of concern.

\begin{figure}
\begin{center}
	\includegraphics[width = \linewidth]{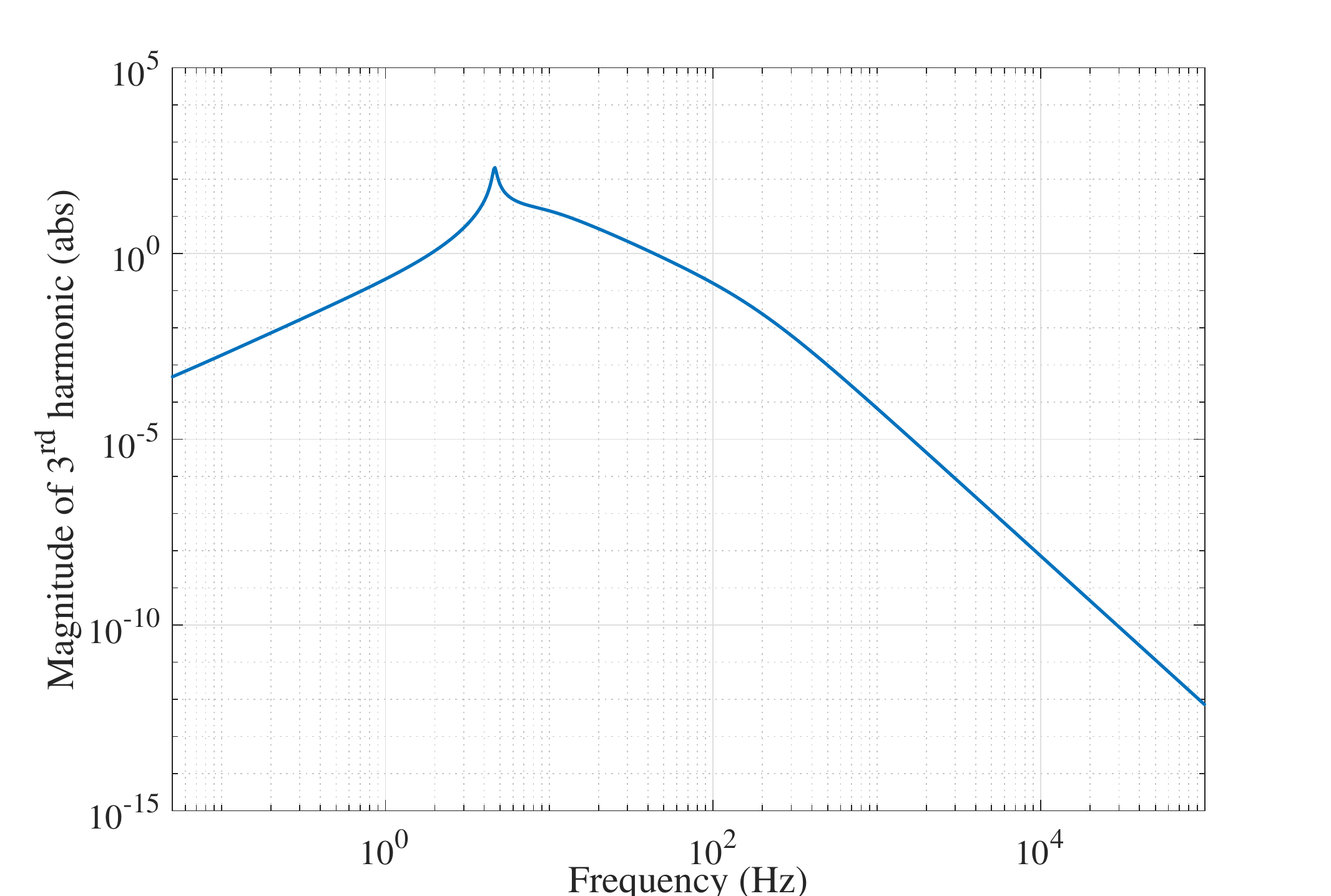} 
	\caption{Third harmonic gain of mass-spring-damper system controlled by CgLp-SORE + PID.}
	\label{fig_h3_sys}
\end{center}
\end{figure}

From (\ref{cglpfore_zero_hosidf_gain}) and (\ref{cglpsore_zero_hosidf_gain}), higher-order harmonic magnitudes are close to zero at low frequencies and reach a constant magnitude asymptotically at high frequencies as given by (\ref{cglpfore_inf_hosidf_gain}) and (\ref{cglpsore_inf_hosidf_gain}). Additionally, from (\ref{cglpfore_zero_hosidf_gain}), (\ref{cglpfore_inf_hosidf_gain}), (\ref{cglpsore_zero_hosidf_gain}), and (\ref{cglpsore_inf_hosidf_gain}), it is clear that there is a trade-off between the magnitudes of higher-order harmonics seen at low and high frequencies depending on the value of $\alpha$ and $\kappa$. However, these values are chosen to satisfy the unity gain of the first harmonic and hence cannot be used to influence the magnitude of higher-order harmonics. However from the context of overall controller design, the harmonics created by CgLp at high frequencies are attenuated by the low pass filter effect of the PID and plant. This can be seen in Figure \ref{fig_h3_sys} which shows the magnitude of third harmonic of a CgLp-GSORE + PID controller designed for a mass-spring-damper system.

\begin{figure}
\begin{center}
	\includegraphics[width = \linewidth]{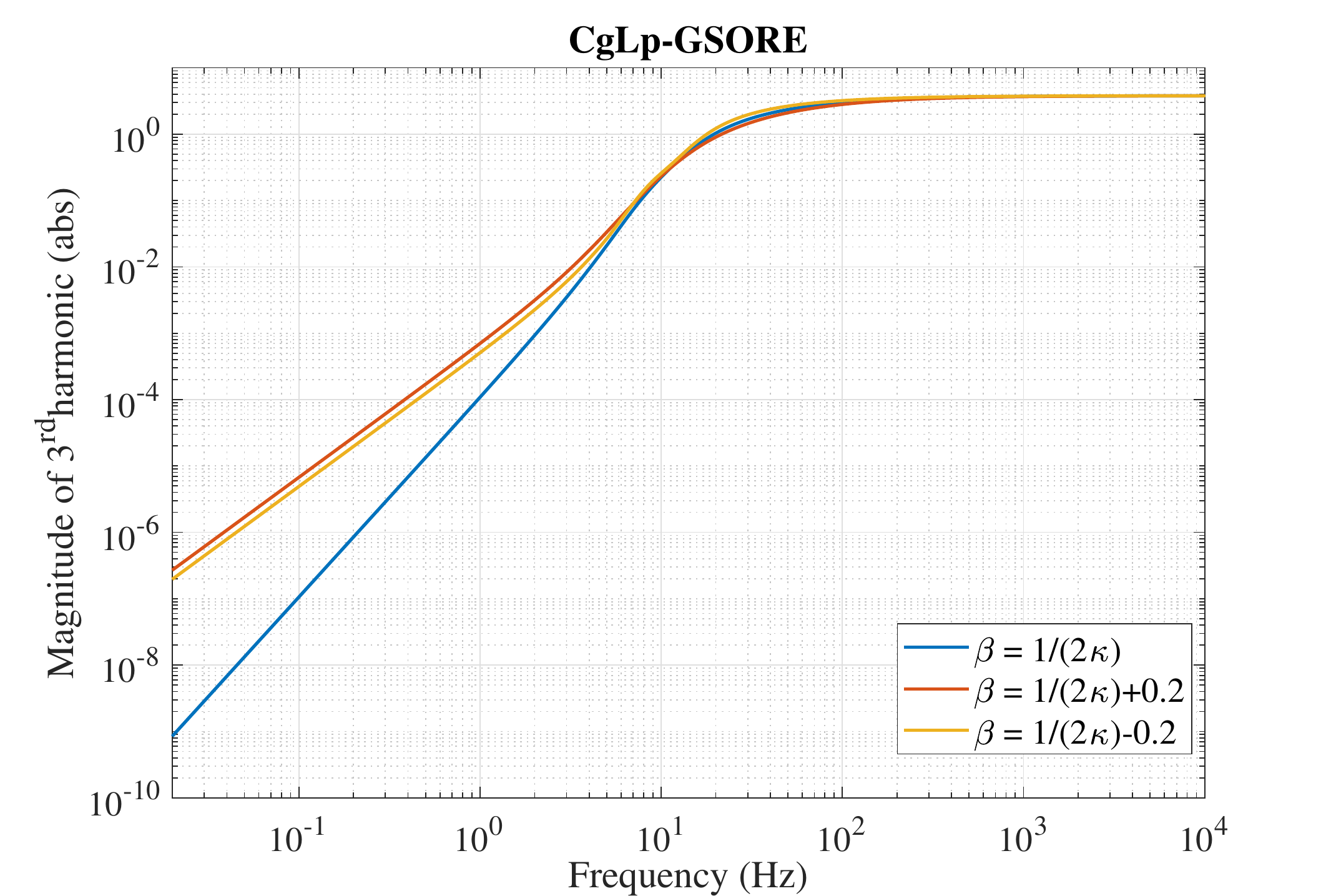} 
	\caption{Third harmonic of CgLp-SORE as influenced by choice of $\beta$ ($\gamma = 0$, $\kappa = \dfrac{1}{\sqrt[4]{1+F^2}}$, $\zeta=1.2$)}
	\label{fig_cglpsore_beta}
\end{center}
\end{figure}

Since higher-order harmonics at high frequencies are attenuated by the plant with additionally discrete controller implementation neglecting all behaviour at really high frequencies, we aim to tune $\gamma$ and $\omega_r$ to obtain the required phase lead with the lowest magnitude of harmonics at low frequencies, i.e., in region of tracking and disturbance rejection. A parameter $\sigma$ is defined in (\ref{sigma-cglpfore}) and (\ref{sigma-cglpsore}) to compare the magnitude for different combinations of  $\omega_r$ and $\gamma$.
\begin{align} 
	\text{CgLp-FORE: }\sigma=\dfrac{1-\gamma}{(\alpha \omega_r)^2} \label{sigma-cglpfore} \\
	\text{CgLp-SORE: }\sigma=\dfrac{1-\gamma}{(\kappa \omega_r)^2} \label{sigma-cglpsore}
\end{align}
In the case of CgLp-SORE, the parameter $\beta$ also has an effect on the magnitude. However it is already established that the lowest higher-order harmonics are achieved when $\beta = \dfrac{1}{2\kappa}$. Since the value of $\zeta$ has no effect on the higher-order harmonics, it is tuned purely to ensure as close to unity gain is achieved by CgLp-SORE. The effect of the choice of $\beta$ on the magnitude of higher-order harmonics is shown in Figure \ref{fig_cglpsore_beta}.

\section{Tuning procedure}
Based on the analysis of the simplified DF and HOSIDF equations, the following procedure is determined and advised for the design of CgLp. 

\begin{enumerate}
	\item For a given amount of phase lead ($\phi$) that has to be provided at the bandwidth by CgLp, calculate the maximum value of $\gamma$ that can be chosen which can at least asymptotically provide the required phase lead. This can be obtained for CgLp-FORE as
	\begin{align*}
	&F = \tan(\phi), \quad
	\gamma_{max} = \dfrac{\dfrac{4}{\pi}-F}{\dfrac{4}{\pi}+F}
	\end{align*}
	However, as noted earlier, due to errors in approximation of the term $e^{(\tfrac{\pi}{\omega} A)}$ at high frequencies, the phase lead as calculated by (\ref{cglpsore_inf_df_phase}) for CgLp-SORE is erroneous and hence the value of $\gamma_{max}$ for CgLp-SORE has to be determined by trial and error or by using the graph provided in \cite{cglp}.
	\item Depending on the value of $\gamma$ determined in the previous step, heuristically choose an array $\Gamma = \{\gamma_1, \gamma_2, ......\}$ where $-1 < \gamma_i < \gamma_{max}$. For all chosen values of $\gamma$, we will design the CgLp to obtain the required phase lead $\phi$ at bandwidth with the minimum magnitude higher-order harmonics in each case. Hence, for each case, follow steps (3) to (6).
  	\item Ensure unity gain at high frequencies by choosing values of $\alpha$ and $\kappa$ according to (\ref{alphachoice}) and (\ref{kappachoice}) respectively.
  	\item For CgLp-SORE, additionally, set $\beta = \dfrac{1}{2\kappa}$ and also choose $\zeta$ to ensure as close to unity gain as possible.
  	\item Determine the value of $\omega_r$ which provides the correct phase lead $\phi$ at the bandwidth. Due to the nonlinear nature of the equations, this has to be achieved through trial and error or a gradient descent algorithm. Additionally, since CgLp achieves a constant phase lead asymptotically, pay attention to ensure that the value of $\omega_r$ determined in each case is as high as possible while simultaneously satisfying the phase lead requirement. This ensures that magnitude of higher-order harmonics at low frequencies is minimised. 
    \item Calculate the factor $\sigma$ and plot it.
    \item Choose the $\gamma$ and $\omega_r$ corresponding to the CgLp design with the lowest $\sigma$. 
    \item Choose a different array $\Gamma$ with the values of $\gamma$ determined by the interval around the value of $\gamma$ chosen in the previous step for further optimisation of choice and repeat steps (3) to (8) till satisfied or based on computation power available.
\end{enumerate}

\section{Practical validation}

\begin{figure}
\begin{center}
	\includegraphics[width=\linewidth]{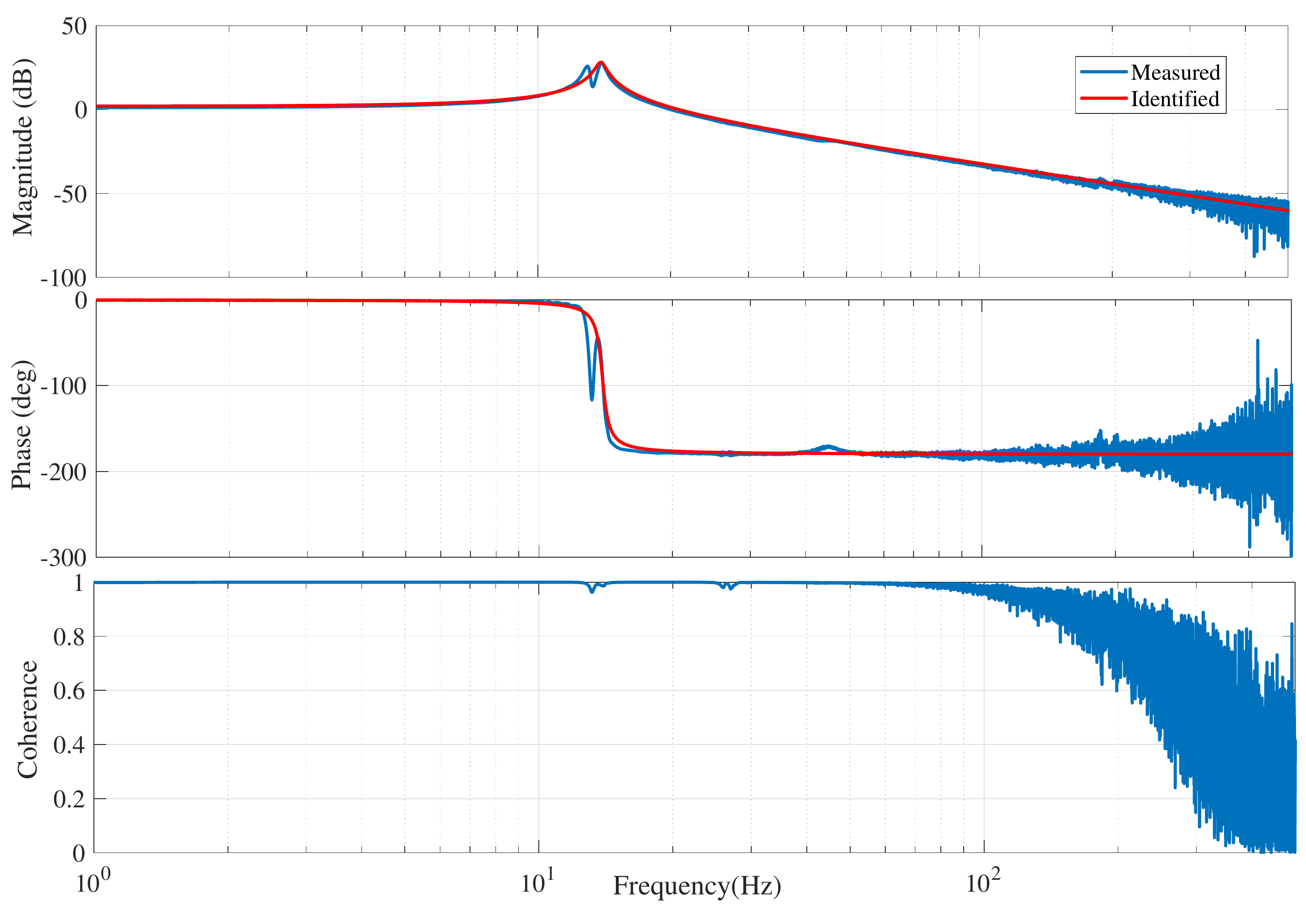} 
	\caption{Frequency response and estimated transfer function of the system}
	\label{fig_system}
\end{center}
\end{figure}

A precision flexure-based positioning stage named `Spyder stage' (Figure) is used for validation. Only one of the actuators (1A) is considered and used for controlling position of mass '3' attached to same actuator resulting in a SISO system. The practically obtained frequency response of the system is shown in Figure \ref{fig_system} and the transfer function is estimated as in (\ref{system}) for stability analysis. Different configurations of CgLp + PID controllers are implemented and the practical results are compared in terms of parameter $\sigma$ for each configuration in order to evaluate the reliability of this parameter and the established procedure of the previous section.
\begin{align} \label{system}
	G(s)=\dfrac{9602.5}{s^2+4.2676s+7627.3}
\end{align}

\begin{figure}
	\begin{center}
		\includegraphics[width = \linewidth]{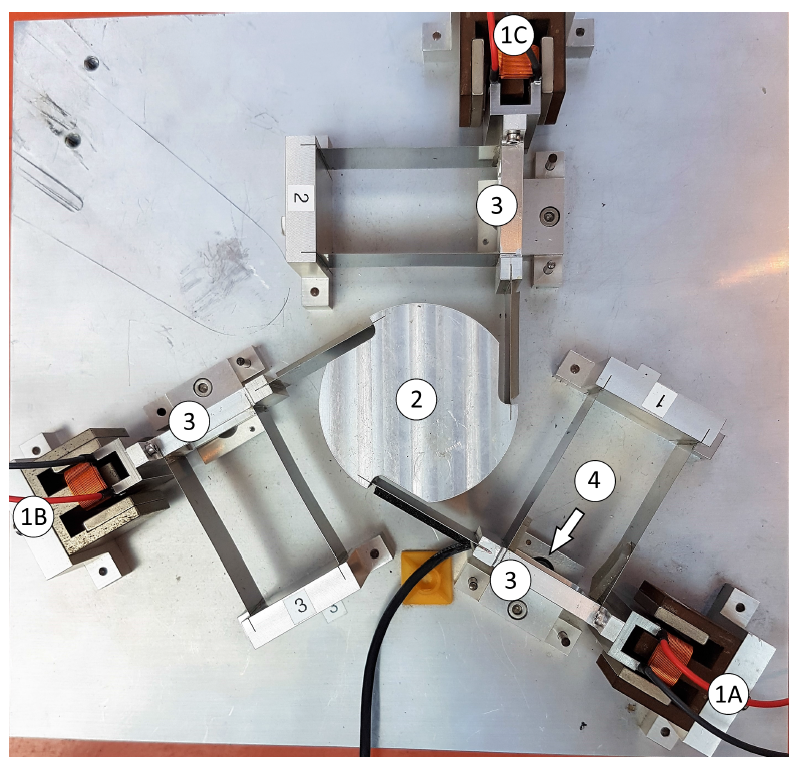} 
		\caption{3 DOF planar precision positioning `Spyder' stage. Voice coil actuators 1A, 1B and 1C control 3 masses (indicated as 3) which are constrained by leaf flexures. The 3 masses are connected to central mass (indicated by 2) through leaf flexures. Linear encoders (indicated by 4) placed under masses '3' provide position feedback with a resolution of $100\ nm$.}
		\label{fig_stage}
	\end{center}
\end{figure}

For all the configurations, the linear part of the controller, i.e., PID is unchanged. Table \ref{general_param} shows the general parameters as applicable to all controllers. Table \ref{fore_hh_table} and Table \ref{sore_hh_table} provide details of CgLp-FOREs and CgLp-SOREs respectively with all CgLp-FOREs providing $40^\circ$ and all CgLp-SOREs providing $60^\circ$ phase lead at the bandwidth ($\omega_c$). 
\begin{table}
\begin{center}
\caption{Parameters of the controllers. $\omega_i$ is the frequency at which integrator action stops and $\omega_f$ is the corner frequency of low-pass filter. All frequencies are in Hz}
\begin{tabular}{ccccccc}
{} & $\omega_c$ & $\omega_i$ & $\omega_f$ & $\alpha$, $\kappa$ & $\beta$ & $\zeta$ \nl 
\hline
CgLp-FORE & 100    & 10     & 500  & $\dfrac{1}{\sqrt{1+F^2}}$ & - & -\nl
CgLp-SORE & 100    & 10     & 500 & $\dfrac{1}{\sqrt[4]{1+F^2}}$ & $\dfrac{1}{2\kappa}$ &1 \dnl
\hline
\end{tabular}
\label{general_param}
\end{center}
\end{table}

\begin{table}
\begin{center}
\caption{Different CgLp-FORE configurations for $40^\circ$ phase lead at the bandwidth with the RMS Error deviation provided for reference frequencies of $1\ Hz$ and $5\ Hz$, where $\omega_r = \omega_c/a$}
\begin{tabular}{ cccccc}
\multirow{2}{*}{Ctrl}  &	\multirow{2}{*}{$\gamma$} &	\multirow{2}{*}{a} &	\multirow{2}{*}{$\sigma$} &	\multicolumn{2}{c}{RMS Error Dev.}\\&&&& $1\ Hz$ & $5\ Hz$ \nl
\hline 
$f_1$ & 0.17    & 7     & 2.33e-04  & 6.4721    & 5.5437 \nl
$f_2$ &   0     & 4.3   & 1.23e-04  & 5.0038    & 4.1356 \nl
$f_3$ & -0.1    & 3     &  8.58e-05 & 4.5575    & 3.5781 \nl
$f_4$ & -0.2    & 2.4   &8.14e-05  & 4.5292    & 3.4315 \nl
$f_5$ & -0.3    & 2     &  8.68E-05 & 4.568     & 3.405 \nl
\hline
\end{tabular}
\label{fore_hh_table}
\end{center}
\end{table}

\begin{table}
\begin{center}
\caption{Different CgLp-SORE setups for $60^\circ$ phase lead at the bandwidth with the RMS Error deviation provided for reference frequencies of $1\ Hz$ and $5\ Hz$, where $\omega_r = \omega_c/a$}
\begin{tabular}{ c c c c c c }
\multirow{2}{*}{Ctrl}  &	\multirow{2}{*}{$\gamma$} &	\multirow{2}{*}{a} &	\multirow{2}{*}{$\sigma$} &	\multicolumn{2}{c}{RMS Error Dev.}\\&&&& $1\ Hz$ & $5\ Hz$ \nl
\hline
$s_1$ &    0.28 &    14    &   2.88e-04 &    16.832 &    15.0511 \nl
$s_2$ &    0.2  &    2.43 &    1.02e-05 &    8.0738 &    7.0878 \nl
$s_3$ &    0.1  &    1.66 &  6.88e-06 & 4.6328 &    3.8989 \nl
$s_4$ &    0    &    1.37 &    7.70e-06 &    5.061  &    5.432 \nl
\hline
\end{tabular}
\label{sore_hh_table}
\end{center}
\end{table}

The main point of validation is to show that the advised procedure ensures that the magnitude of higher-order harmonics are reduced and minimal for the optimal configuration and hence this should have the least deviation between expected error and measured error among all the configurations. The expected RMS error can be calculated as 
\begin{align}
	\text{RMS Error} = \dfrac{SF(\omega) \cdot Ref(\omega)}{\sqrt{2}} \label{DF_rms_error}
	\end{align}
where SF is sensitivity function calculated based on DF and Ref is the amplitude of reference signal to the system. The deviation is calculated as the ratio of measured RMS error to expected RMS error with a smaller value indicating better prediction and a corresponding smaller error seen in practice.

\begin{figure}
\begin{center}
	\includegraphics[width=\linewidth]{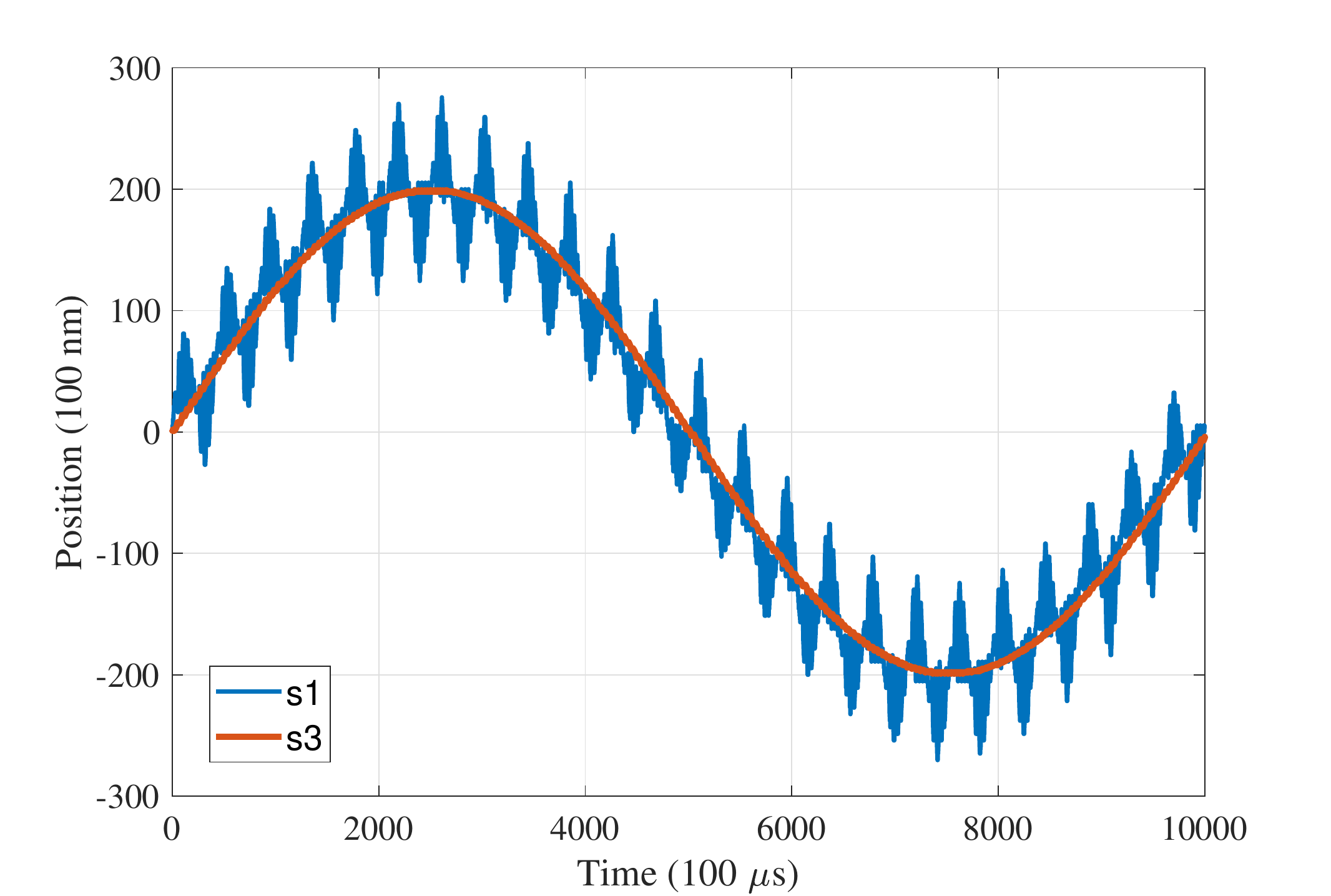} 
	\caption{Comparison between the practical result of the controllers $s_1$ and $s_3$. The reference is a sinusoidal wave with amplitude of 20 microns and 1 Hz. }
	\label{s1_s3_practical}
\end{center}
\end{figure}

The value of $\sigma$ is an indication of the magnitude of the higher-order harmonics and hence a prediction measure for optimal CgLp configuration. Based on the values listed in Table \ref{fore_hh_table} and Table \ref{sore_hh_table}, $f_4$ and $s_3$ have least magnitude harmonics and hence are expected to have best tracking performance. While this is true for a reference frequency of $1\ Hz$, this is not true for CgLp-FORE for $5\ Hz$. However, the difference between the values of both $\sigma$ and RMS error deviation are small for $f_4$ and $f_5$ and hence can be considered as an aberration. In table \ref{sore_hh_table}, the trend of $\sigma$ matches well with the RMS error deviation. Figure \ref{s1_s3_practical} compares the practical tracking results of the controllers $s_1$ and $s_3$ which produce the highest and the least magnitude of higher-order harmonics among the chosen controllers. The stark difference in tracking performance for the controllers can be clearly seen and showcases the need to systematically design these reset controllers.

\section{Conclusions}
The high precision and bandwidth requirements of the high-tech industry are pushing the limits of linear controllers which are restricted by water-bed effect. Reset controllers are a promising alternative which can be easily adopted into the PID framework with the CgLp element shown to provide significant performance improvements. However, most work in literature for tuning of reset controllers is based on describing function analysis. This can be highly inaccurate especially for precision applications and additionally the various choices in CgLp design which can provide different closed-loop performance is not captured by DF. Hence, in this paper, we present an analysis based approach to using the higher-order sinusoidal input describing function (HOSIDFs) which provide information related to the higher-order harmonics and develop a simple tuning procedure to design optimal CgLp based PID controllers. The iterative procedure is outlined and used to design different controllers for a precision positioning stage. The practical results validate the presented procedure and showcases that reset controllers especially based on CgLp design are industry ready.

\bibliography{Bibl} 
\end{document}